# Volumetric incorporation of NV diamond emitters in nanostructured F2 glass magneto-optical fiber probes

Adam Filipkowski,[1,2] Mariusz Mrózek,[3] Grzegorz Stępniewski,[1,2] Tanvi Karpate,[1] Maciej Głowacki,[4] Mateusz Ficek,[4] Wojciech Gawlik,[3] Ryszard Buczyński,[1,2] Adam Wojciechowski,[3] Robert Bogdanowicz,[4] and Mariusz Klimczak[1,*]

[1] Faculty of Physics, University of Warsaw, Pasteura 5, Warsaw, 02-093, Poland

[2] Łukasiewicz Research Network - Institute of Microelectronics and Photonics, Al. Lotników 32/46, Warsaw, 02-668, Poland

[3] Institute of Physics, Jagiellonian University in Kraków, Łojasiewicza 11, Kraków, 30-348 Poland

[4] Faculty of Electronics, Telecommunications and Informatics, Gdańsk University of Technology, Narutowicza 11/12, Gdańsk, 80-233 Poland

*Email: mariusz.klimczak@fuw.edu.pl

**Abstract**

Integration of optically-active nanodiamonds with glass fibers is a powerful method of scaling of diamond's magnetic sensing functionality. We propose a novel approach for integration of nanodiamonds containing nitrogen-vacancy centers directly into the fiber core. The core is fabricated using nanostructurization, i.e. by stacking the preform from 790 soft glass canes, drawn from a single rod dip-coated with nanodiamonds suspended in isopropyl alcohol. This enables manual control over distribution of nanoscale features, here – the nanodiamonds across and along the fiber core. We verify this by mapping the nanodiamond distribution in the core using confocal microscopy. The nanodiamonds are separated longitudinally either by 15 μm or 24 μm, while in the transverse plane separation of approximately 1 μm is observed, corresponding to the individual cane diameter in the final fiber, without significant agglomeration. Filtered, red fluorescence is observed with naked eye uniformly along the fiber. Its magnetic sensitivity is confirmed by in optically detected magnetic resonance recorded with a coiled, 60-cm-long fiber sample with readout contrast of 1.3% limited by microwave antenna coverage. NV fluorescence intensity in 0 to 35 mT magnetic field is also demonstrated, allowing magnetometry applications with a large B-field dynamic range in absence of microwaves.

**Keywords:** nanodiamonds, nitrogen vacancy centers, magnetic sensing, optical fibers,

## 1. Introduction

Sensing functionalities of diamond encompass electrical, magnetic, and dielectric interaction with matter and can be shaped by chemical modification and size of the diamond crystals [1]. Doping using plasma-enhanced chemical vapor deposition (CVD) has opened means for modification of diamond's refractive index and conductivity [2]. The latter facilitates the use of diamond in electrochemical sensors as an ultrawide bandgap semiconductor with electron mobility exceeding 2000 $cm^2$/Vs [3,4]. Incorporation of nitrogen-vacancy color centers (NV), which unfolds novel magnetic properties, can be realized by either CVD, or other methods, such as graphite-to-diamond phase transition under high pressure and high temperature (the HPHT method) or the detonation of diamond crystals mixed with $C_aH_bN_cO_d$ type compounds [5-7].

Practical sensor devices based on NV-doped diamond require integration with structures that would leverage on the scaling up of these different sensing modalities. For example, the pick and drop method can be used to transfer crystalline diamond particles from growth substrates onto various flexible platforms, which enabled overcoming the intrinsic brittleness of diamond in strain sensors sensitive for strains below 0.1 [8]. Optical readout of diamond's magnetic response in the optically detected magnetic resonance (ODMR) technique motivates the integration of diamond particles with optical fibers. Implementations involving fiber tip deposition or various evanescent field realizations like fiber tapers are intensively investigated with nano- (NDs) or microdiamonds (MDs) containing NV or other color centers [9-11]. Fiber tip functionalization with color center nanodiamonds has been demonstrated in micrometer-scale spatial resolution, non-contact thermal mapping application with 20 mK/√Hz thermal resolution [12]. Prior designs of photonic crystal fiber tip functionalization with NV nanodiamonds were demonstrated in the context of optical adjustment-free single photon sources [13]. These results revealed impressive NV fluorescence collection efficiency with fibers, which was comparable to far-field imaging through an objective with a numerical aperture larger than 0.8. NV fluorescence collection efficiency in the range of 35-37% was reported for tapered single-mode fibers, in which the tapered section was in direct contact with a diamond micro-waveguide for possible single photon emitters and efficient quantum optics interfaces [10,14]. These implementations relate to ultra-sensitive, highly localized interaction of NV emitters with the surrounding. Fluorescence collection of NV centers in diamond over extremely distributed sites in the tens-of-meters scale was also demonstrated using a fiberized setup [15], however, in this implementation the NV fluorescence was collected by photodiode chips distributed along 90 m length of a fiber capillary, in which a droplet with a single microdiamond particle was pushed along by applied air pressure.

Volume incorporation of diamond micro- and nano-particles into a glass and subsequent drawing of optical fibers has been proposed as a radically alternative method of scaling of NV emission yield [16,17]. Although theoretically straightforward, it is hindered with oxidation and graphitization of diamond particles during the high temperature processing steps, including glass melting and fiber drawing at a tower. The choice of glass hosts is thus limited to low melting point soft glasses and indeed the first demonstrated diamond particle volume-doped glass fibers were drawn using tellurite glasses with melting (diamond incorporation) and drawing temperatures of 610°C and 400°C, respectively [17,18]. It has to be noted, that tellurite glass fibers are tedious in handling, especially to non-experts in specialty fiber development, due to the poor mechanical properties of the glass itself. Alternatively, silicate soft glasses can be used, like the F2 lead-silicate glass (Schott), which albeit requiring a higher drawing temperature around 700°C, possesses superior mechanical robustness compared to tellurite glasses [19]. Thermal decomposition or oxidation of diamond particles are not the sole concerns when the volume incorporation of optical fibers is considered. Melting of bulk soft glass from raw materials mixed with diamond particles, followed by drawing of the cast and extruded glass into fiber preforms, results in an inhomogeneous distribution of the NDs across the fiber core and along the fiber length. Such random distribution contributes to the scattering losses and reduces NV fluorescence guiding efficiency, thus work on the means of optimizing coupling of this fluorescence to the fiber guided modes have already begun [20].

In this work, we report on a novel volume incorporation approach for the integration of nitrogen vacancy centers containing nanodiamond particles with optical fibers. A step-index fiber is developed by stacking the core preform from 790 canes made of lead-silicate glass, which after drawing at a fiber drawing tower, constitute a solid fiber core. It is a simplification of the core nanostructurization method, based on the effective medium theory, in which a stack of glass canes with sub-wavelength individual diameters (after the final thinning at a fiber drawing tower) with slightly different optical properties and arranged in a designed pattern, yields certain averaged ("effective") distribution of a specific property, for example, the refractive index profile [21]. Here, the nanostructurization concept is simplified by using the same type of glass canes across the stack. The canes have been drawn from a single glass rod, dip-coated with crystalline nanodiamond particles containing NV centers suspended in isopropyl alcohol, similarly to the method proposed earlier [14]. The diamond-coated cane stack was inserted in a tube made of thermally matched silicate glass and drawn at a fiber drawing tower into a final multimode, step-index fiber with 50 μm core diameter and 125 μm outer diameter. Mapping of physical distribution of the NV emitters embedded in the core has been performed using confocal microscopy imaging, revealing remarkably even distribution of single crystalline diamonds across the transverse and longitudinal planes of the fiber core. The developed fiber was tested in two experiments involving its magnetic sensitivity stemming from the presence of NV nanodiamonds in its core. In one scenario the fiber was spun on a microwave antenna and optically detected magnetic resonance was measured with readout contrast of up to 1.3%, which was scalable with and limited by the coverage of the fiber by the antenna. In the second test experiment, we measured NV fluorescence intensity change in a 35 mT dynamics range of magnetic flux density from an electromagnet.

## 2. Experimental methods

### 2.1. Nanodiamond-incorporated optical fiber fabrication

Fiber development procedure begun with the preparation of preform components, which included the cladding tube and a solid glass rod for fabrication of nanodiamond-coated glass canes constituting the fiber core stack. F2 lead silicate glass (Schott) was selected as the base fiber glass and specifically, it was used for nanodiamond functionalization and core stacking. Nominally, the lead silicate F2 glass contains between 40 and 50 weight % each of silica and lead oxide, which are complemented by up to 10% each of potassium oxide and sodium oxide and below 1% of arsenic trioxide. Its refractive index is $n_1 = 1.6199$ at the wavelength of 589.3 nm. A modified composition within the above limits was melted in-house in the form of a tube, with the (linear) refractive index $n_2 = 1.6133$ at the wavelength of 589.3 nm. This glass was used for the cladding tube of the fiber preform. An index-guiding fiber was thus possible to obtain from these materials.

The actual fabrication procedure was split into four steps, which are schematically shown in Fig. 1a. The first involved integration of nanodiamonds with the glass rod. We adapted the dip-coating approach reported earlier by D. Bai et al. [19]. The nanodiamonds used for coating were suspended in isopropanol, with the diamond concentration of 0.05%. We used nanodiamonds with an average particle size of 750 nm (MDNV1um, Adámas Nanotechnologies), which is a compromise between fluorescence yield and scattering loss introduced by the particles in the core. Before dip-coating, a cylindrical cuvette with the nanodiamond suspension was stirred in an ultrasonic cleaner for 20 minutes and then was moved onto a magnetic stirrer for 3 minutes. The cuvette was then placed again in an ultrasonic cleaner for 5 minutes, and again for 3 minutes on the magnetic stirrer. Then the cuvette was placed in the ultrasonic cleaner for the final 5 minutes. After this, the F2 glass rod (diameter of 30 mm) was slowly immersed in the nanodiamond suspension in the cuvette. After lifting from the cuvette, the excess liquid was allowed to drain back into the cylinder and residual isopropanol to evaporate from the surface of the glass rod. This coating procedure was repeated 10 times.

In the second step, the dip-coated F2 glass rod was drawn into canes. To ensure that no contaminants were present in the fiber the coated glass rod was placed in the drawing tower furnace and then heated up to 400°C in a dry oxygen atmosphere. This temperature is high enough to burn most small contaminates on the glass, while leaving the nanodiamonds intact. This was verified before dip-coating by thermogravimetric analysis (TGA) of the diamond powder, which indicated mass change related to oxidation at temperatures exceeding 550°C. F2 glass fiber drawing requires a higher temperature of around 700-750°C, which is unacceptable for diamond in an oxygen atmosphere. However, TGA reveals that diamond mass change of only 0.5% occurs at this temperature range when the oxygen atmosphere is replaced with nitrogen. A nitrogen atmosphere was thus introduced to the furnace and the temperature was increased to 740°C. After the start of the drawing process of the nanodiamond-coated rod, the temperature was reduced to 710°C and the rod was drawn at that temperature. The rod was drawn from 30 mm diameter down to 0.5 mm ± 0.02 mm canes. In the third step, the drawn canes were cut to 12 cm length and stacked inside the in-house prepared glass tube with an external diameter of $d_e = 40$ mm and an internal diameter of $d_i = 16$ mm. 790 glass canes were fitted inside forming the structured glass preform. The preform was then drawn at the drawing tower from the initial diameter down to a 4 mm diameter structured rod. During this process, the nitrogen atmosphere was again introduced into the furnace. Additionally, after the drawing process had started, low vacuum was introduced selectively to the core stack to ensure integration of the individual nanodiamond-coated canes. This was crucial during this part of the drawing, as rough diamond coating of the individual rods traps atmospheric gasses between canes and leads to uneven glass integration. In the final step the structured rod with a fully fused, nanodiamond-doped core was drawn into the final fiber. The external diameter of the fabricated fiber is $d_f = 125$ μm (excluding the protective acrylic coating), while the diameter of the core is $d_c = 50$ µm. During the drawing, the diameter of the fully fused canes in the core stack was decreased to $d_r = 1.5$ µm. The scanning electron microscope image of the fiber is shown in Fig. 1b. Due to low refractive index contrast, the core area is indistinctive from the cladding, although it can be well seen under a standard optical microscope, which is shown in Fig. 1c. The mode field of the red fluorescence of NV$^-$ color centers, filtered with a red high-pass filter to cut out the 532 nm laser pump, is shown in Fig. 1d.

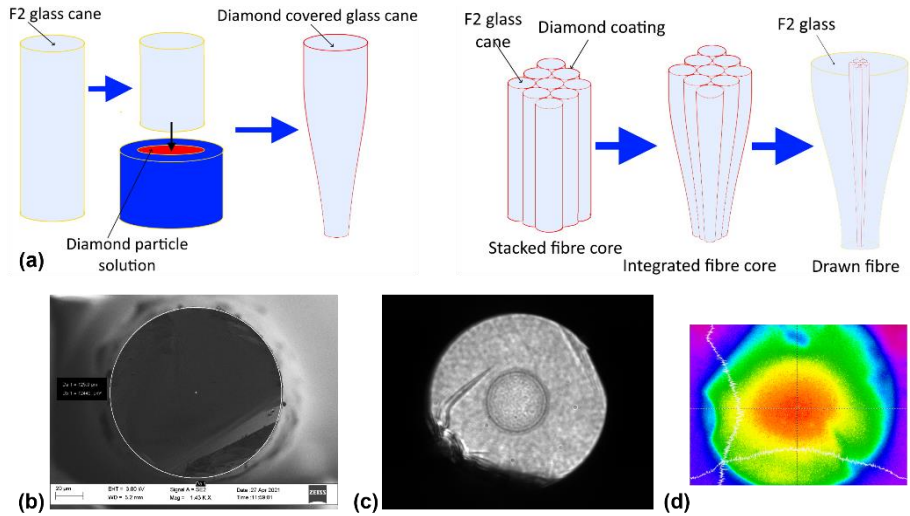

**Fig. 1.** Nanodiamond-doped nanostructured core fiber: (a) fiber development procedure, (b) scanning electron microscope image of the fiber, (c) optical microscope image of the fiber, (d) mode field image of the fiber output over a red high-pass filter.

## 3. Results and discussion
### 3.1. Nanodiamond fluorescence and spatial distribution in the fiber

Core stacking using multiple canes employed here is an approach derived from the effective medium theory from arbitrary refractive index shaping in fibers and micro-optics components [21]. In essence, the approach relies on the possibility to manually determine the nanoscale distribution of specific components of the fiber structure. Although simplified here and not aimed at refractive index control specifically, the approach used in this work is expected to retain its principal advantage of enabling manual control over the distribution of nanoscale features, here – the nanodiamond particles in the fiber core. In the preceding part, we have demonstrated that one of the key challenges in nanodiamond-doped fiber drawing from dip-coated glass preforms involving successful integration of preform glass components in presence of a nanodiamond layer, can be solved by selective application of vacuum to the core area. The obtained, fully fused, nanodiamond-doped core fiber has also been demonstrated to couple NV color center fluorescence into its guided modes. This part of work moves on to reporting results of spatial, i.e. the transverse and longitudinal mapping of nanodiamonds distribution in the fiber core.

Characterization of the distribution of diamond particles in the developed F2 glass fiber (and the fiber's sensing potential described in the following section) was carried out using the confocal microscopy technique. Three-dimensional scans were recorded with a commercial confocal microscope system (Zeiss LSM710). A 4 mm long section of fiber was placed on a microscope slide surrounded with immersion oil (n = 1.52). Fig. 2a schematically shows the nanodiamond mapping arrangement with the principal planes of imaging. The propagation direction along the fiber was in the Y axis, while the transverse plane of the fiber was located at the XZ plane. A 3-D image viewed from the YZ plane – shown in Fig. 2c, i.e. along the propagation direction along the fiber, was acquired by stacking 137 individual images in the XY plane with a 0.5 µm step along the Z axis. The image height is around 68 µm, which just slightly exceeds the core diameter of 50 µm. The length of the imaged fiber section, i.e. the length of the fiber, was limited to around 354 µm, which stemmed from the size of the field of view of the microscope system. The bright spots correspond to nanodiamonds and their spatial distortion in the image is caused by astigmatism due to the difference of around Δn = 0.09 between the fiber refractive index the surrounding immersion oil. The transverse, XZ plane of the fiber is shown in Fig. 2b. The core is outlined with a yellow circle. The image shows the transverse distribution of nanodiamonds corresponding to the longitudinal distribution shown in Fig. 2c over 354 μm length of the fiber. Both the longitudinal and transverse plane images enable distinguishing single nanodiamond particles in the volume of the core without significant agglomerations despite one, which is manifested as a brighter spot, located around 80 μm from the fiber sample facet and slightly offset from the core center.

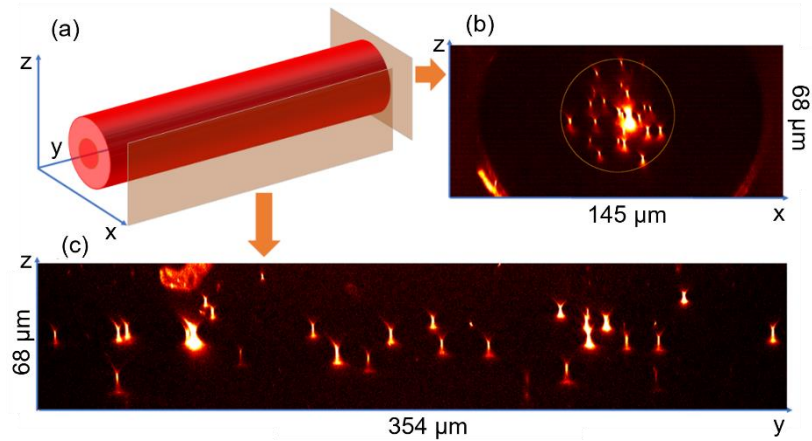

**Fig. 2.** Confocal microscope scan of the developed fiber: (a) schematics of the fiber sample with the imaging planes marked, (b) nanodiamond fluorescence recorded for the XZ plane, and (c) the YZ plane.

To evaluate the nanodiamond distribution in the fiber core in a quantitative manner, we measured the distances between the immediately (directly) adjacent nanodiamond pairs along the optical fiber and also across its core. Fig. 3 shows histograms obtained this way for two cases. Histograms were computed by projecting the image (signal intensity) onto one axis, i.e. the longitudinal histogram was obtained from projection of signal intensity onto the Y axis, and the transverse distribution histogram was calculated from signal intensity projection onto the X axis. Two arguments support the representative character of this data, despite the short fiber sample imaged in the confocal microscope system. Firstly, the drawing procedure and constant parameters (i.e. feed and pull rate of the rod, canes and fiber preform) facilitate a fiber with nanodiamonds distributed with similar statistics along the entire drawn length. Secondly, the concentration of diamonds in the developed fiber, due to multiple diamond-coated cane stacking in the core, exceeds significantly that in the diamond ring F2 glass fiber or the tellurite glass fibers drawn from glass billets melted from raw materials mixed with diamond powders, reported in previous demonstrations [17-19]. Thus even at the short section of the sample under the confocal microscope, the nanodiamond concentration provided a meaningful sample for statistics. In the longitudinal plane of the fiber, two major fractions can be distinguished, as shown in the histogram in Fig. 3a. The nanodiamond particles are separated either by 15 μm or 24 μm. In the transverse plane, nanodiamond separation of roughly 1 μm dominates, as shown in the histogram in Fig. 2b, which corresponds to the 1.5 μm diameter of the estimated individual cane diameter in the final fiber core. Multiple cane stacking has an important advantage in the context of this characteristics because replacing nanodiamond-coated canes with pure glass canes in the stack enables manipulation of the transverse distribution and provides means of e.g. controlling nanodiamond concentration at specific areas of the fiber core. Table 1 contains a summary of the recently demonstrated magneto-optic fiber probes with nanodiamonds incorporated into the volume of the fiber core, either by direct melt doping, or by the dip-coating technique.

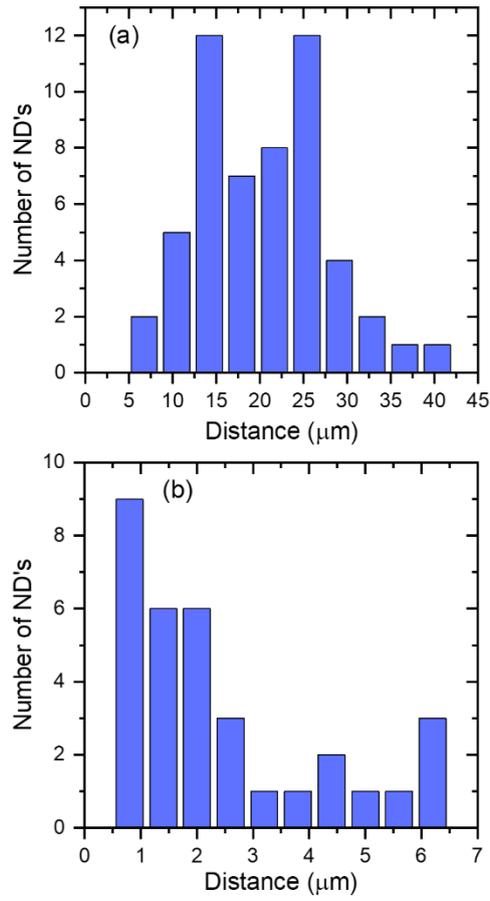

**Fig. 3.** Histograms of distribution between individual, adjacent nanodiamonds: (a) along the optical fiber (Y axis) and: (b) in the transverse plane the optical fiber (X axis).

**Table 1.** Summary of fabrication, structural and magnetic sensing characteristics of realized magneto-optic fiber probes with ND-doped cores.

| Characteristic | Tellurite glass step index fibers [17,18] | F2 glass fibers with nanodiamond ring [20] | F2 glass nanostructured ND-doped fibers [this work] |
|---|---|---|---|
| Diamond particle size | 45 nm | 1 µm | 750 nm |
| Temperature during ND incorporation into glass | 610 °C | 660-690 °C | 710 °C |
| Fiber drawing temperature | 400 °C | 660-690 °C | 710 °C |
| Core/cladding refractive indices | Single glass, either 1.98 or 2.00 (at λ = 1064 nm) | Single glass (F2), 1.6199 (λ = 589.3 nm) | 1.6199 (core) and 1.6133 (cladding), λ = 589.3 nm |
| ND incorporation technique | ND powder added to glass melt | Dip-coating of fiber core preform in ND suspension, followed by drawing | Dip-coating of high-index glass rods in ND suspension, followed by drawing and stacking of multiple canes into a structured core preform |
| ND distribution – longitudinal | Random, further diluted during drawing | Random, further diluted during drawing | Semi-random, 14 µm and 25 µm – dominant distances between adjacent ND pairs, longitudinal dilution suppressed by cane stacking |
| ND distribution – transverse | Random | Determined by circular shape of the dip-coated core preform, randomized along core circumference due to dilution during drawing | Determined by cane diameter in final fiber core, 1 µm dominant distance between adjacent ND pairs, controllable by manual arrangement of cane stack layout |
| Demonstrated magnetic field sensing performance (detailed discussion in Section 4) | ODMR read-out contrast of 3.5% (point excitation of fiber) | ODMR read-out contrast of 2.5% (point excitation of fiber) | ODMR read-out contrast of 1.3% (limited by antenna coverage) MW-free B-field measurement with 35 mT dynamic range (>20 cm long section of fiber excited) |

### 3.2. Proof-of-principle fiber magnetometry

In the final part of the work, the developed fiber was evaluated for its applicability in magnetic sensing. Two types of experiments were carried out, both involving a 60 cm long fiber sample. The first one was measurement of the optically detected magnetic resonance (ODMR), followed by measurement of the dependence of ND fluorescence intensity on the magnetic field without presence of microwaves. We begin with the ODMR experiment and the setup used in this part is shown in Fig. 4a. Optical readout of the fiber's magnetic response was realized under excitation from a 532 nm continuous-wave laser (Sprout-H). The green light was coupled into the fiber from both ends: on one end a standard collimating lens was used, and on the other light was coupled through a dichroic mirror (Thorlabs DMLP567) and a microscope objective (Olympus UPLFLN 40×, NA = 0.75). NV nanodiamond fluorescence was separated from the green light at the dichroic mirror and residual 532 nm signal was filtered out using a 600 nm high-pass filter (FEL0600, Thorlabs). The NV fluorescence was detected with a Si avalanche photodetector (APD130A, Thorlabs). This setup allowed effective detection of fluorescence over a wavelength range of 600–850 nm, as well as comparing the excitation efficiency from either end of the fiber sample. In order to observe the optically detected magnetic resonance (ODMR), a microwave (MW) oscillating field at the frequency around 2.87 GHz was generated using a signal generator (SRS, SG386) and a high-power amplifier (Mini-Circuits ZHL-16W-43+) connected to a loop-gap type antenna structure on a printed circuit board [22]. One fiber end was passed through the opening in the center of the antenna loop which allowed the interaction of this fiber section with the MW field. Fluorescence and ODMR signals were then observed on an oscilloscope. Fig. 4b and Fig. 4c show a section of the fiber during measurements under 532 nm excitation: without and with the 600 nm high-pass filter. Red fluorescence was observable with the naked eye (through the high-pass filter) uniformly along the entire fiber.

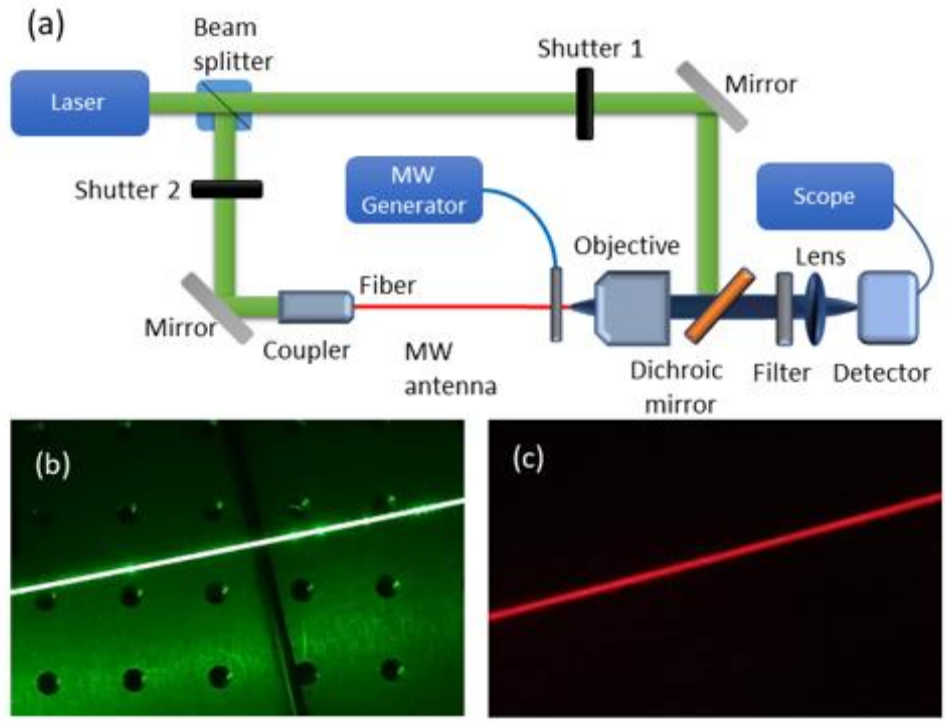

**Fig. 4.** (a) Experimental setup for magnetic sensing experiments with the developed fiber, (b) standard camera image of a short fiber section under 532 nm excitation on an optical table and (c) the same fiber section seen through a red high-pass filter (FEL0600, Thorlabs).

Optical fluorescence spectra of the investigated NV- fiber samples were recorded using a compact grating spectrometer (AvaSpec-3648-USB2, Avantes). The spectra, shown in Fig. 5, were collected in two configurations. In the first one, excitation and detection took place from the same end of the fiber – this was realized with shutter 1 open, Fig. 4a. In the second measurement, the excitation and detection occurred from the opposite ends of the fiber – shutter 2 was opened, Fig. 4a. The zero-phonon line (ZPL) is marked with a dotted line. The recorded signal was normalized to the intensity at the ZPL wavelength. The difference in the two spectra is assigned to scattering occurring during propagation in the fiber, when excitation and detection of NV color centers fluorescence occur at opposite ends of the fiber.

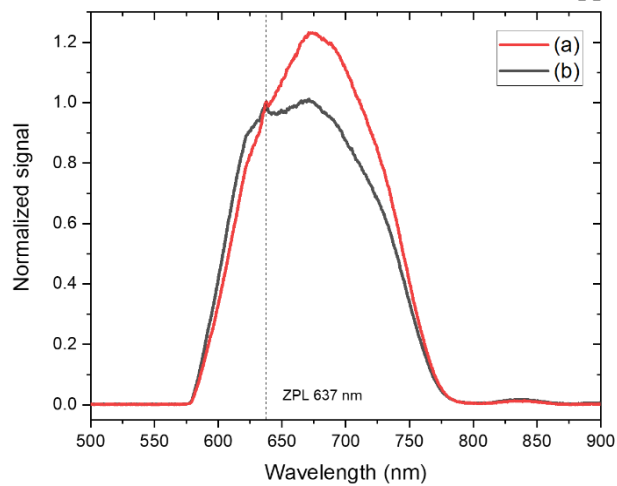

**Fig. 5.** Fluorescence spectrum for two cases: (a) excitation and detection from the same end of the fiber, (b) excitation and detection from opposite ends of the fiber.

Out of the different sensing modalities accessible with NV color centers in diamond and microwaves, including measurements of magnetic field and temperature, here we realized a simple case of ODMR-based magnetic field sensing. ODMR readout was carried out in two configurations with the fluorescence excitation and detection taking place either for the same fiber end, or from the opposite fiber ends. Three magnetic flux densities were consecutively set in the measurements: 0 G, 15 G, and 21 G. The observed spectra are shown in Fig. 6. For the 0 G setting, the recorded ODMR spectrum did not differ significantly from the single-crystalline diamond, while in non-zero fields significant differences were recorded. The spectra are strongly inhomogeneously broadened by the random orientation of nanocrystals with respect to the direction of the magnetic field [22]. The outermost edges of the spectra correspond to the nanodiamonds with the maximum projections of the magnetic field, i.e., those oriented along the field direction. An approximately linear increase of the spectrum width with the field strength can also be observed. The main difference between the spectra recorded from either the same or the opposite end to the excitation direction is the level of readout contrast. This is because the antenna was located approximately 5 mm from one end of the fiber, so that in configuration with excitation and detection from the same fiber end, not all NV- color centers are excited by the microwave field. It is to be noted, that the readout contrast could be further improved with antenna design to increase coverage of the fiber by the MW field.

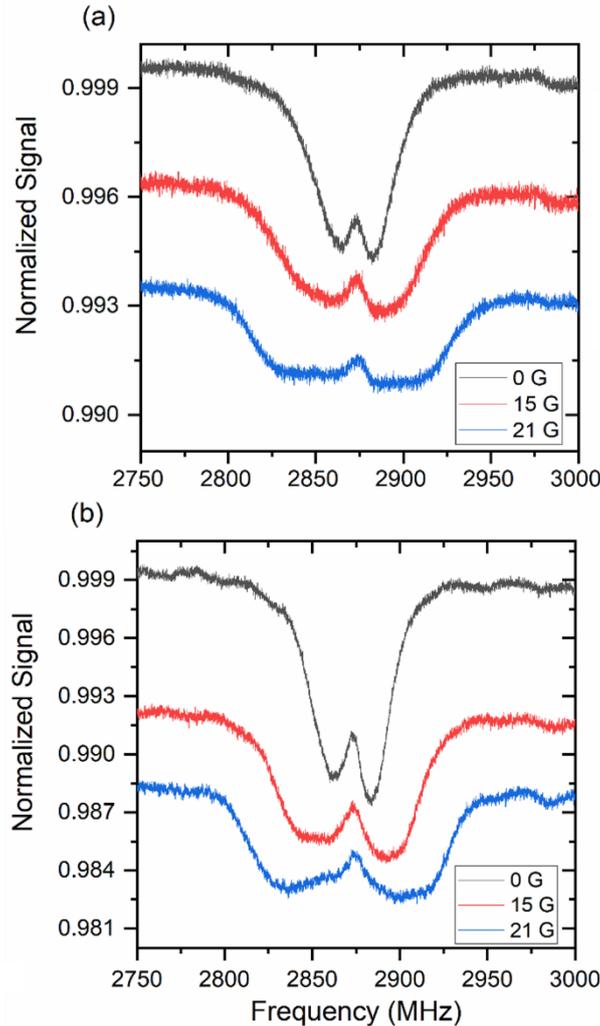

**Fig. 6.** ODMR spectrum collected in two configurations: (a) excitation and detection from the same fiber end, (b) excitation and detection from opposite fiber ends.

NV- magnetometry can also be performed with the developed fiber in a MW-free arrangement. Fig. 7 shows the photoluminescence measured as a function of the magnetic field. The inset shows the diagram of the experiment. To expose the entire optical fiber to interaction with the magnetic field from the neodymium magnet, the fiber was coiled into an approximately 3 cm diameter loop, as shown in a photograph in the inset of Fig. 7 (the magnet is also visible above the fiber coil). The initial gradual decrease in photoluminescence is associated with a reduction in emission of the non-aligned NV$^-$ centers due to spin-mixing [23]. This decrease in fluorescence can be used as a proof of concept magnetometry experiment with a 35 mT dynamic range.

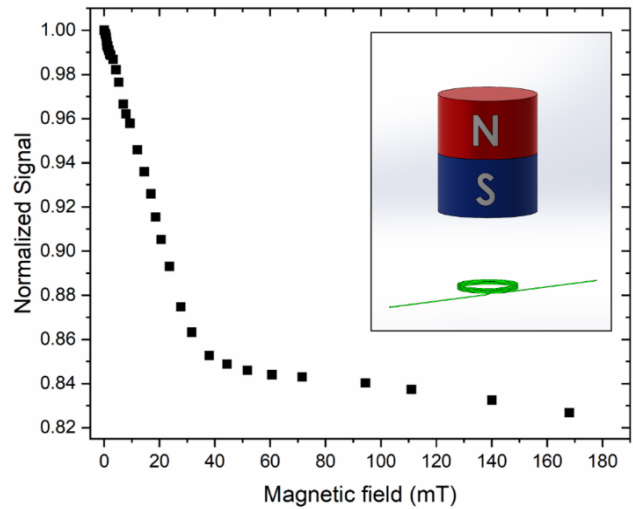

**Fig. 7.** Nitrogen-vacancy photoluminescence as a function of the applied magnetic field normalized to the photoluminescence at 0 mT. The inset shows the diagram of the experiment.

## 4. Conclusions
The main highlight of the work is achieving an even, controllable distribution of single nanodiamond particles across and along a step-index optical fiber core. Nanodiamonds distribution in the transverse and longitudinal planes of the fiber were mapped in the volume of the core along the length of 0.7 mm using confocal microscopy. Two main fractions in the longitudinal nanodiamond separation were around 15 µm and 24 µm, and transverse separation was typically 1-3 µm, which corresponds with the estimated final (drawn fiber) 1.5 µm diameter of canes constituting the pre-final drawing core stack. In contrast to previously reported implementations involving drawing of soft glass fibers directly from an ND-suspension dip-coated glass preforms, stacking of multiple ND-coated (and thinned at drawing temperature at a drawing tower) canes, combined with carefully adjusted drawing conditions, the fiber demonstrated in this work had fully fused core without airholes neither at the core-cladding nor at the intra-core interfaces. In addition to that, no evidence of significant agglomeration has been observed, despite the uniform distribution of nanodiamonds in the transverse plane of the core.

The demonstrated stacking fabrication approach does not hinder the magnetic sensing functionality of the developed proof-of-concept fiber. We have verified this by recording ODMR spectra, in which the readout contrast (around 1.3%) was limited by the design of the microwave antenna used. However, it can be easily enhanced, e.g. using a microwave resonator which can effectively excite a centimeter-sized loop of optical fiber [24]. Additionally, we have demonstrated a microwave-free detection mode, in which fluorescence level can be mapped to the magnetic field value in a broad dynamic range of 0-35 mT.

The core of the proof-of-concept fiber was stacked in an approach derived from the core nanostructurization concept based on the effective medium theory [21]. This method has been successfully implemented in various arbitrary refractive index profile shaping realizations in an optical fiber, for example for developing of a graded-index, mid-infrared fiber made from CVD-incompatible chalcogenide glasses or for fiber-tip integrable, double-flat surface, graded-index microlenses [25,26].

In the context of NV-fiber magnetometry, core nanostructurization carries two advantages. Firstly, it enables virtually unrestricted positioning of nanodiamonds relative to the transverse plane of the core for optimal coupling of the NV$^-$ fluorescence to the fiber guided modes. A recent theoretical study of this issue indicated significant coupling improvement by localization of the NV$^-$ close to the cladding-core boundary[20]. Secondly, here the separation between the diamond particles can be controlled directly, by adjusting the dipping suspension concentration, for example implementing core stack designs involving diamond-functionalized canes arbitrarily interleaved with pure glass canes. This methodology enables manual control over the transverse layout of the NDs, extending the existing models with practical ways of optimizing NV⁻ fluorescence coupling to the guided modes in optical fiber-based ODMR and magnetometry probes.

**Acknowledgements**

This work was supported by Foundation for Polish Science grant
TEAM NET POIR.04.04.00-00-1644/18. All authors contributed equally to this work.

**Declaration of competing interest**

The authors declare that they have no known competing financial interests or personal relationships that could have appeared to influence the work reported in this paper.